%% file: paper.tex
\documentclass[10pt,conference]{IEEEtran}
\usepackage[table]{xcolor}
\usepackage[utf8]{inputenc}
\usepackage{amsmath}
\usepackage{graphicx}
\usepackage{siunitx}
\usepackage{todonotes}
\usepackage{array}
\usepackage[frozencache]{minted}
\usepackage{soulutf8}
\usepackage{tikz}
\usepackage{subcaption}
\usepackage{cite}
\usepackage{float}
\usepackage{balance}
\usepackage{url}

\usetikzlibrary{shapes}
\usetikzlibrary{patterns, fit}
\usepackage{adjustbox}

\setminted[java]{
xleftmargin=20pt,
framesep=2mm,
linenos=true,
breaklines=true,
tabsize=4,
encoding=utf8,
fontsize=\footnotesize,
frame=none,
fontsize=\scriptsize,
escapeinside=||
}

\usepackage{xspace}
\usepackage{tcolorbox}
\usepackage{enumitem}
\usepackage{multirow}

\newcommand{\highlight}[1]{\begin{tcolorbox}[leftrule=1mm,rightrule=1mm,toprule=0mm,bottomrule=0mm,left=2pt,right=2pt,top=1pt,bottom=1pt]
#1
\end{tcolorbox}
}

\newboolean{showcomments}
\setboolean{showcomments}{true}
\ifthenelse{\boolean{showcomments}}
 { \newcommand{\mynote}[2]{
      \fbox{\bfseries\sffamily\scriptsize#1}
        {\small$\blacktriangleright$\textsf{\emph{#2}}$\blacktriangleleft$}}}
        { \newcommand{\mynote}[2]{}}

\makeatletter
\newcommand{\ostar}{\mathbin{\mathpalette\make@circled\star}}
\newcommand{\make@circled}[2]{%
  \ooalign{$\m@th#1\smallbigcirc{#1}$\cr\hidewidth$\m@th#1#2$\hidewidth\cr}%
}
\newcommand{\smallbigcirc}[1]{%
  \vcenter{\hbox{\scalebox{0.97778}{$\m@th#1\bigcirc$}}}%
}
\makeatother

\newcolumntype{g}{>{\columncolor{gray!30}}l}
\newcolumntype{h}{>{\columncolor{gray!30}}c}
\newcommand{\raicc}[0]{\textsc{RAICC}\xspace}
\newcommand{\iccta}[0]{\textsc{IccTA}}
\newcommand{\ic}[0]{\textsc{IC3}}
\newcommand{\droidbench}[0]{\textsc{DroidBench}}

\begin{document}

\hyphenation{Pend-ing-Intents}
\hyphenation{Pend-ing-Intent}

\title{RAICC: Revealing Atypical Inter-Component Communication in Android Apps}

\author{\IEEEauthorblockN{Jordan Samhi\IEEEauthorrefmark{1},
Alexandre Bartel\IEEEauthorrefmark{1}\IEEEauthorrefmark{2}, Tegawendé F. Bissyandé\IEEEauthorrefmark{1} and
Jacques Klein\IEEEauthorrefmark{1}}
\IEEEauthorblockA{\IEEEauthorrefmark{1} SnT,
University of Luxembourg, firstname.lastname@uni.lu}
\IEEEauthorblockA{\IEEEauthorrefmark{2} DIKU,
University of Copenhagen, ab@di.ku.dk}}

\maketitle

\begin{abstract}
\input{sections/abstract}
\end{abstract}

\begin{IEEEkeywords}
     Static Analysis, Android Security
\end{IEEEkeywords}

\input{sections/introduction}
\input{sections/icc_challenges}

\input{sections/motivation}
\input{sections/approach}

\input{sections/evaluation}
\input{sections/limitations}

\input{sections/related_work}
\input{sections/conclusion}

\input{sections/data_availability}
\input{sections/acknowledgment}

%\balance
\bibliographystyle{IEEEtran}
\bibliography{bib}

\end{document}

%% file: sections/abstract.tex
Inter-Component Communication (ICC) is a key mechanism in Android.
It enables developers to compose rich functionalities and explore reuse within and across apps.
Unfortunately, as reported by a large body of literature, ICC is rather ``complex and largely unconstrained'', leaving room to a lack of precision in apps modeling.
To address the challenge of tracking ICCs within apps, state of the art static approaches such as \textsc{Epicc}, \iccta{} and \textsc{Amandroid} have focused on the documented framework ICC methods (e.g., startActivity) to build their approaches.
In this work we show that ICC models inferred in these state of the art tools may actually be incomplete: the framework provides other atypical ways of performing ICCs.
To address this limitation in the state of the art, we propose \raicc{} a static approach for modeling new ICC links and thus boosting previous analysis tasks such as 
ICC vulnerability detection, 
privacy leaks detection, 
malware detection, etc. 
We have evaluated \raicc{} on 20 benchmark apps, demonstrating that it improves the precision and recall of uncovered leaks in state of the art tools.
We have also performed a large empirical investigation showing that Atypical ICC methods are largely used in Android apps, although not necessarily for data transfer. We also show that \raicc{} increases the number of ICC links found by 61.6\% on a dataset of real-world malicious apps, and that \raicc{} enables the detection of new ICC vulnerabilities. 

%% file: sections/introduction.tex
\section{Introduction}
\label{introduction}

Android apps heavily rely on the \emph{Inter-component communication} (ICC) mechanism to implement a variety of interactions such as sharing data~\cite{chin2011analyzing}, triggering the switch between UI components or asynchronously controlling the execution of background tasks. Given its importance, the research community has taken a particular interest in ICC, reporting on various studies that show how ICC can be exploited in malicious scenarios: ICC can be leveraged to easily connect  malicious payload to a benign app~\cite{li2017-Piggybacking}, leak private data~\cite{gordon2015information,wei2014amandroid,li2015iccta}, or perform app collusion~\cite{abro2017android}. These scenarios are generally executed by passing \texttt{Intent} objects, which carry the data and information about explicitly/implicitly targeted components~\cite{xu2016iccdetector}. 
Tracking information across Intents to link components that may be connected via ICC thus becomes an important challenge for the analysis of Android apps.

The resolution of ICC links (identification of the source and target components, type of the components, etc.) is a well-studied topic in the literature. 
Approaches such as \textsc{EPICC}~\cite{octeau2013effective}, \textsc{COAL/IC3}~\cite{octeau2015composite}, SPARTA~\cite{barros2015static} or DroidRa~\cite{li2016droidra} have contributed with analysis building blocks in this respect. 
The ICC links (also called ICC models) generated by these tools are key and even mandatory for several Android app analysis tasks. 
(1) In the case of data flow analysis, ICC poses an important challenge in the community: ICC indeed introduces a discontinuity in the flow of the analysis, since there is no direct call to the target component life-cycle methods in the super-graph (aggregation of control flow graphs~\cite{allen1970control} of caller and callee methods in the absence of a single Main method). 
Several tool-supported approaches such as  \textsc{Amandroid}~\cite{wei2014amandroid}, \textsc{IccTA}~\cite{li2015iccta} and \textsc{DroidSafe}~\cite{gordon2015information} have been proposed in the literature to cope with this issue. 
To overcome the discontinuity in the flow of the analysis, all these three tools rely on an inferred ICC model to identify the target component and the ICC methods in order to artificially connect components. 
(2) In the case of Android malware detection, a tool such as  \textsc{ICCDetector}~\cite{xu2016iccdetector} leverages the ICC model generated by \textsc{EPICC} to derive ICC specific features that are used to produce a Machine-Learning model in order to detect new type of Android malware. 
(3) In the case of vulnerability detection, \textsc{EPICC} leverages its own ICC model to detect ICC vulnerabilities, defined in~\cite{octeau2013effective} as the sending an Intent that may be intercepted by a malicious component, or when legitimate app components, --e.g., a component sending sms messages-- are activated via malicious \textit{Intent}.

In all these cases, the proposed tools rely on a comprehensive modeling of the ICC links. 
However, a major limitation in ICC resolution relates to the fact that state of the art approaches consider only well-documented ICC methods such as \texttt{startActivity()}. Indeed, we have discovered that several methods from the Android framework can also be used to implement ICC although the official Android documentation does not specifically discuss it~\cite{services, fundamentals, intents}.
Actually, ICC can also be performed by leveraging Android objects (e.g., \texttt{PendingIntent} or \texttt{IntentSender}) that have been little studied in the literature, and through framework methods that can atypically be used to launch other components.

We have initially observed an atypical ICC implementation during the manual reverse engineering of an Android app
that we identified as part of research on logic bomb detection.
This app uses the method \texttt{set(int, long, Pending\-Intent)} of the \texttt{AlarmManager} class for triggering a \texttt{BroadcastReceiver} which in turn is used to launch a \texttt{Service} component. 
Such an implementation appeared suspicious since it seems artificially complex: it is possible to directly call the \texttt{sendBroadcast} method instead of leveraging an \texttt{Alarm\-Manager}. 
We further performed extensive investigations and found that several dozens of methods of the Android framework can atypically start a component with objects of type \texttt{Pending\-Intent} and/or \texttt{IntentSender}.
We use the term "atypical" to reflect the fact that, according to the method definitions, their role is not primarily to start a component (as ICC methods typically do) but to perform some action (e.g., set an alarm or send an SMS). 
Unfortunately, with such possibilities,  an attacker could rely on such methods to perform ICC-related malicious actions. Existing state-of-the-art approaches, because they do not account for atypical methods in their models, would miss detecting such ICC links.

Our work explores the prevalence of \emph{Atypical ICC} (AICC) methods in the Android framework as well as their usages in Android apps. We then propose an approach for resolving those AICC methods and an instrumentation-based framework to support state-of-the-art tools in their analysis of ICC. 

In summary, we present the following contributions:
\begin{itemize}
    \item We present findings of a large empirical study on the use of AICC methods in malicious and benign apps.
    \item We propose a tool-supported approach named \raicc
 for resolving AICC methods using code instrumentations in order to generate a new APK with standard ICC methods. 
We demonstrate that this instrumentation boosts state of the art tools in various Android analysis tasks.
    \item We improve \droidbench{}~\cite{droidbench} with 20 new apps using AICC methods for assessing data leak detection tools.
\end{itemize}

The rest of the paper is organized as follows.
First, we present how state-of-the-art performs with ICC in Section~\ref{background}.
Then, in Section~\ref{motivation}, we give an example and explain why we are studying atypical inter-component communication.
In Section~\ref{approach}, we detail \raicc{}, our tool-supported approach.
We evaluate \raicc{} and present our results in Section~\ref{evaluation}.
In section~\ref{limitations}, we present the limitations of the approach.
Finally, we discuss the related-work in Section~\ref{related_work} and conclude in Section~\ref{conclusion}.

%% file: sections/icc_challenges.tex
\section{How do state of the art Analyzers handle ICC?}
\label{background}

Android apps are composed of components that are bridged together through the ICC mechanism. 
The \texttt{Activity} component implements the UI visible to users while {\em Service} components run background tasks and {\em Content Provider} components expose shared databases. 
An app may also include a {\em Broadcast Receiver} component to be notified of system events. 
The \emph{Manifest} file generally enumerates these components with the relevant permission requests.

Components are activated by calling relevant ICC methods provided in the Android framework. These ICC methods are also used to pass data through an {\em Intent} object, which may explicitly target a specific component or may implicitly refer to all components that have been declared (through {\em Intent Filters}) capable of performing the Intent actions.

The ICC mechanism challenges static analysis of apps.
Indeed, consider Listing~\ref{code_icc_example} in which the \texttt{MainActivity} component
launches the \texttt{TargetActivity} component.
The discontinuity in the control-flow is clear since there is no direct method call between \texttt{Main\-Activity} and \texttt{Target\-Acti\-vity}. 
Off-the-shelf Java static analyzers that analyze normal method calls would not be able to detect the link between 
the ICC method \texttt{startActivity} and the \texttt{Target\-Ac\-ti\-vity} component.
Hence, if a data flow analysis is performed, none of the data-flow values can be propagated correctly.
This is since ICC methods trigger internal Android system mechanisms which redirect the call to the specified component.

\begin{listing}
\begin{minted}{java}
public class MainActivity extends Activity {
  protected void onCreate(Bundle b) { ...
    Intent i = new Intent(this, TargetActivity.class);
    this.startActivity(i);
  }
}
public class TargetActivity extends Activity 
  { protected void onCreate(Bundle b) {} }
\end{minted}
    \caption{An example of how ICC is performed between two components.}
    \label{code_icc_example}
\end{listing}

Therefore, Android static analyzers have to preprocess the application in order to add explicit method calls. 
That is what state-of-the-art tools like \iccta{}~\cite{li2015iccta}, \textsc{DroidSafe}~\cite{gordon2015information} and \textsc{AmanDroid}~\cite{wei2014amandroid} do with different techniques.
If we take the example of \iccta{}, it first relies on IC3~\cite{octeau2015composite} to infer the ICC links. Among other information, IC3 identifies the ICC methods (e.g., \texttt{startActivity} in Listing~\ref{code_icc_example}) and resolves the target components (e.g., \texttt{Target\-Ac\-ti\-vity} in Listing~\ref{code_icc_example}). Then,  \iccta{} replaces any ICC method call with a direct method call that passes the correct \texttt{Intent}. 
Thus, the discontinuity disappears and the link to the target component is directly available in the super-graph (see Figure 3 of \iccta{} paper~\cite{li2015iccta}).
The idea that we reuse in this paper is the code instrumentation that allows preprocessing an app for constructing the missing links to be processed by any analysis.

Nevertheless, in this paper, we will see that state-of-the-art approaches only rely on well-documented methods for performing inter-component communication.
We aim at improving their precision by revealing previously un-modeled ICC links.

%% file: sections/motivation.tex
\vspace{-2mm}
\section{Atypical ICC Methods}
\label{motivation}

Static analysis of Android applications is challenging due to 
the specificity of the Android system's inter-component communication (ICC) mechanism.
Therefore, as we have overviewed in Section~\ref{background}, researchers have to come up with approaches for considering and resolving ICC.
In this section, we show that one developer can perform atypical ICC by taking advantage of specific methods of the Android framework.

We define an \emph{atypical ICC method} (AICC method) as a method 
allowing to perform an inter-component communication while it is not its primary purpose. 
These AICC methods rely on \texttt{Pending{\-}Intents} and \texttt{IntentSenders}.
\texttt{PendingIntents} objects are wrappers for \texttt{Intents}.
They can only be generated from existing \texttt{Intents} and describe those latter.
They can be passed to different components and especially to different applications.
When doing so, the receiving app is granted the right to perform the action described in the \texttt{Pending{\-}Intent} with the same permissions and identity of the source app.
This introduces a security threat in which a component could perform an action for which it does not have the permission but it is granted this latter through the \texttt{PendingIntent}.
This security threat has been studied by Gro{\ss} et al.~\cite{gross2018pianalyzer}.
An important fact is that \texttt{PendingIntents} are maintained by the system and represent a copy of the original data used to create it. The \texttt{PendingIntents} can thus still be used if the original app is killed.
IntentSenders objects are encapsulated into \texttt{Pending\-Intents}.
They can be retrieved from a \texttt{PendingIntent} object via the method \texttt{getIntentSender()}.
Basically, they can be used the same way than \texttt{PendingIntents} and represent the same artifact.

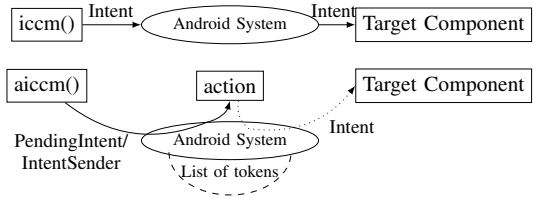
\begin{figure}[h]
    \begin{adjustbox}{width=0.80\columnwidth,center}
        \input{tikz/icc_representation.tikz}
    \end{adjustbox}
    \vspace{-3mm}
    \caption{Difference between normal ICC method and AICC method. {\footnotesize Tokens represent \texttt{PendingIntents} and \texttt{IntentSenders}. Action represents the primary purpose of the AICC method (e.g. send an SMS). An action might influence the list of tokens in the Android system, which will later process the list and send \texttt{Intents}. The dotted line indicates that the triggering of the target component may depend on the result of an action.}}
    \label{icc_representation}
\end{figure}

The abstract representation of AICC methods is shown in Figure~\ref{icc_representation}.
The upper part of the figure shows how standard ICC
methods behave.
They communicate with the Android system via \texttt{Intents} to execute another component.
The lower part represents how AICC methods behave.
They perform the action they are meant to do through the Android system and at the same time the \texttt{Pen\-dingIn\-tent} or the \texttt{IntentSender} is registered in a token list in the Android system~\cite{pendingintent,intentsender}.
The action may or may not influence the decision for the system to launch the component, depending on the AICC method.
For example, a \texttt{PendingIntent} could only be launched in case of the success of the action.
Also, the Android system can receive a cancellation of a token from the app. (e.g., cancel an alarm). 
In that case, the target component would not be launched.

The tokens represent the original data used for generating a \texttt{PendingIntent} or an \texttt{IntentSender}.
It means that if the application modifies the \texttt{Intent} used to construct the \texttt{PendingIntent}, it does not affect the token as it is a copy of the original data.
More importantly, if the application is killed, the list is maintained in the Android Framework and the components can still be executed.

%%%%%%%%%%%%%%%%%%%%%%%%%%%%%%%%%%%%%%%%%%%%%%%%%%%%%%%%%%%%%%%%%%%%%%%%%%%%%%%%%%%%%%%%%%%%%%%%%%%%%%%%%%%%%%
%%%%%%%%%%%%%%%%%%%%%%%%%%%%%%%%%%%%%%%%%%%%%%%%%%%%%%%%%%%%%%%%%%%%%%%%%%%%%%%%%%%%%%%%%%%%%%%%%%%%%%%%%%%%%%

\noindent
\textbf{A concrete Example:} %\subsection{AlarmManager}\label{subsec:alarm}
As described in Section~\ref{introduction}, while manually analyzing a malicious application, we noticed that it used the \texttt{AlarmManager} for performing ICC.
The interesting piece of code of this malicious app is presented in Listing~\ref{code_malware}.
We can see a \texttt{PendingIntent} created from an \texttt{Intent} targeting the component \texttt{AlarmListener}. 
The latter simply launches the \texttt{Service} component responsible for retrieving external commands via HTTP.
For launching the class \texttt{AlarmListener}, the developer could have used the method \texttt{sendBroadcast} (\texttt{Alarm\-Listener} extends \texttt{Broad\-cast\-Re\-cei\-ver}), but instead it used the AICC method \texttt{set(int, long, PendingIntent)}.

\begin{listing}[h]
\begin{minted}{java}
public void reconnectionAttempts() {
  Calendar cal = Calendar.getInstance();
  cal.add(12, this.elapsedTime);
  Intent i = new Intent(this, AlarmListener.class);
  intent.putExtra("alarm_message", "Wake up Dude !");
  PendingIntent pi = PendingIntent.getBroadcast(this, 0, i, 134217728);
  AlarmManager am = (AlarmManager) getSystemService("alarm");
  am.set(0, cal.getTimeInMillis(), pi);
}
\end{minted}
    \vspace{-2mm}
    \caption{A simplified example of how the method \texttt{set} of the \texttt{AlarmManager} class is used in a malware.}
    \label{code_malware}
\end{listing}

When we focus more on the way \texttt{PendingIntent} works, we understand why the developer used this technique.
Indeed, in this example, the alarm is set up to go off after 5, 10, or 30 minutes.
But what happens if the user closes the app before it goes off?
In fact, the alarm will go off anyway and execute the target component.
This is due to the fact that when setting an alarm, the \texttt{PendingIntent} is maintained by the Android system until it goes off or gets canceled.
We can see the power of such a method to perform ICC.
It could be used in different scenarios by an attacker to perform its malicious activities.

Furthermore, AICC methods carry information in \texttt{Intent} objects that are also embedded in \texttt{PendingIntent} or \texttt{IntentSender} objects.
Therefore, they can carry different types of information, leading to potential sensitive data leaks. Our benchmark includes examples scenarios for such leaks. 

%% file: tikz/icc_representation.tikz
\begin{tikzpicture}
    %\draw (0,0.8) node[] {ICC Method:};
    \node[draw] (m1) at (-4,0) {iccm()};
    \node[draw,ellipse,scale=.8] (as) at (-1,0) {Android System};
    \node[draw] (target_component) at (2.5,0) {Target Component};

    \node[draw] (m2) at (-4,-1) {aiccm()};
    \node[draw] (action) at (-1,-1) {action};
    \node[draw,ellipse,scale=.8] (as1) at (-1,-1.9) {Android System};
    \node[draw] (target_component1) at (2.5,-1) {Target Component};
    \draw[dashed] (-2,-2.09) arc[start angle=180,end angle=360,x radius=1,y radius=.7];
    \draw (-1,-2.4) node[] {\footnotesize List of tokens};

    \draw[->,>=latex] (m1) -- (as);
    \draw[->,>=latex] (as) -- (target_component);
    %\draw[->,>=latex] (m2) -- (action);
    %\draw[->,>=latex] (as1) -- (action);
    \draw[->,>=latex] (m2) to[bend right] (as1.140) to[bend right] (action.south);
    \draw[->,>=latex,dotted] (action.300) to[bend right] (-0.7,-1.7) to[bend right] (target_component1.west);
    %\draw[->,>=latex,dotted] (as1) -- (target_component1.west);
    %\draw[->,>=latex,dotted] (as1.352) to[bend right] (target_component1.south);

    \draw (-2.95,0) node[above]{\small{Intent}};
    \draw (0.7,0) node[above]{\small{Intent}};
    \draw (-3.6,-2.2) node[above]{\small{PendingIntent/}};
    \draw (-3.6,-2.5) node[above]{\small{IntentSender}};
    \draw (1,-1.9) node[above]{\small{Intent}};
\end{tikzpicture}

%% file: sections/approach.tex
\section{Approach}
\label{approach}

In this paper, we aim at resolving those AICC methods through app instrumentation~\cite{arzt2013instrumenting}.
The goal for the new app is to be analyzable by state-of-the-art Android static analyzers. We first introduce in section~\ref{approach_list_indirect_icc_methods} how we gather a comprehensive list of AICC methods.
Then, in section~\ref{approach_tool_overview} we describe how we leveraged this list of methods to improve the detection of inter-component communications leading to the increase of precision metrics of existing Android-specific static analyzers.

\subsection{List of Atypical ICC Methods}
\label{approach_list_indirect_icc_methods}

As explained in previous sections, during the reverse-engineering of Android applications, we stumbled upon a malicious app making the use of the \texttt{set()} method of the \texttt{AlarmManager} class with a \texttt{PendingIntent} as parameter to stealthy perform an ICC (in this case, to start a \texttt{BroadcastReceiver}).
Thanks to this example, we realized that 
(1) \texttt{Intent} and method such as \texttt{startActivity} are not the only main starting points of ICC, 
(2) other objects (e.g. \texttt{PendingIntent}) and other methods (e.g. \texttt{AlarmManager.set()}) can play a similar role.

Motivated by this discovery, we were eager to check if this atypical mechanism is restricted to this \texttt{set()} method and this \texttt{PendingIntent} object. 
In other words, are there other atypical methods in the Android framework? Are there other classes such as \texttt{PendingIntent}?
To answer these questions, we performed a comprehensive analysis of the Android framework. 

We retrieved from the Android framework, from SDK version 3 to 29 (versions 1 and 2 being unavailable), all the methods that take as a parameter an object of type \texttt{PendingIntent}.
We obtained a list of 163 unique methods.
The next step was to manually analyze all of them in order to only keep those allowing to perform ICC.
The list reduced to 85 methods, indeed some methods have a \texttt{PendingIntent} as a parameter but cannot perform ICC (e.g., \texttt{android.bluetooth.le.Bluetooth\-LeScanner.stopScan(PendingIntent)}).

To identify classes similar to \texttt{PendingIntent}, we followed a simple heuristic. 
We search for all class names containing the string $Intent$. 
This search yielded 19 classes that we manually checked. 
Finally, we identified one new class, \texttt{IntentSender}, which, according to the Android documentation, has the same purpose as \texttt{PendingIntent}.
We scanned again the Android framework to retrieve all the methods that take as a parameter an object of type \texttt{IntentSender}, and we discovered 17 new methods for performing atypical inter-component communication.

To improve the confidence in our list of AICC methods, we performed further analyses. 
In particular, we downloaded the source code of Android and studied the implementation of some of the AICC methods we gathered.
This approach aimed at finding patterns that we used to find similar usage in the Android framework, we assumed that other AICC methods use the same patterns. 
We also made some assumptions, e.g., considering the subclasses of those studied.
Unfortunately, we were not able to uncover additional AICC methods.

At this stage, our list reached a length of $102$ ($85+17$) methods.
It was all without counting the 9 methods of \texttt{PendingIntent} and \texttt{Intent\-Sender} classes that directly allow launching a component.
For example, the \texttt{send()} method of the \texttt{PendingIntent} class allows to directly communicate with a targeted component, likewise for method \texttt{sendIntent()} of class \texttt{IntentSender}.
Finally, our list reached 111 methods.

\begin{listing}
\begin{minted}{java}
public class MainActivity extends Activity {
  @Override
  protected void onCreate(Bundle b) {...
    Intent i = new Intent(this, TargetActivity.class);
    TelephonyManager tm = (TelephonyManager) getSystemService(Context.TELEPHONY_SERVICE);
    String imei = tm.getDeviceId();
    i.putExtra("SensitiveData", imei);
    PendingIntent pi = PendingIntent.getActivity(this, 0, i, 0);
    
    (1.a) pi.|\colorbox{yellow}{send}|();

    (2.a) IntentSender s = pi.getIntentSender();}
    (2.b) s.|\colorbox{yellow}{sendIntent}|(this, 0, null, null, null);

    (3.a) AlarmManager am = (AlarmManager) getSystemService("alarm");
    (3.b) am.|\colorbox{yellow}{setExact}|(0, System.currentTimeMillis() - 100, pi);

    (4.a) LocationManager l = (LocationManager) getSystemService("location");
    (4.b) l.|\colorbox{yellow}{requestLocationUpdates}|(0, 0, new Criteria(), pi);
    }
}
\end{minted}
    \caption{Examples of how {AICC methods} (in \colorbox{yellow}{yellow}) can be used to perform inter-component communication. }
    \label{code_iiccm_examples}
\end{listing}

In Listing~\ref{code_iiccm_examples} we illustrate the usage of four AICC methods (chosen for their brevity).
On the first lines (4-8) objects necessary to the AICC methods are instantiated.
An \texttt{Intent} is instantiated at line 4.
At lines 5-6, a sensitive information, the device unique identifier, is retrieved and stored in the \texttt{imei} variable.
At line 7, the IMEI is added as an extra information in the intent.
At line 8, the \texttt{PendingIntent} is instantiated with the intent containing the IMEI.
Then, from line 10, we present four ways 
of launching the \texttt{TargetActivity} component through AICC methods.

\highlight{ We gathered a comprehensive list of 111 methods, called AICC methods, allowing to perform atypical inter-component communication.}

\subsection{Tool Design}
\label{approach_tool_overview}

\begin{figure}[h]
\centering
    \input{tikz/overview.tikz}
    \caption{Overview of our open-source tool \raicc{}.}
    \label{overview}
\end{figure}
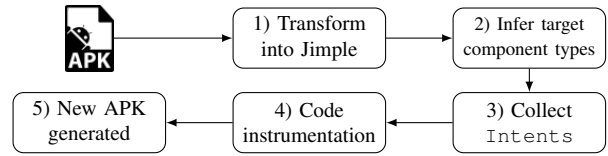

\textbf{General Idea:}
The overview of our open-source tool called \raicc is depicted in Figure~\ref{overview}.
The general idea is to instrument a given Android app to boost it by making it aware of ICC links.
For instance, if a  \texttt{PendingIntent} is used with an AICC method to start an activity, \raicc will instrument the  app's source code by adding a method  \texttt{startActivity()} with the right intent as parameter. 
This method is added at a point of interest in the app, i.e., just after the AICC method call. 
To perform this instrumentation, \raicc needs 
(1) to infer the possible values/targets of ICC objects (e.g., \texttt{Intent}); %at program point of interests 
(2) resolve the type of the target component in order to instrument with the right standard ICC methods (e.g., \texttt{startActivity()} if the target component is an Activity, \texttt{startService()} if the target component is a Service, etc.).

\vspace{-2mm}
\vspace{-2mm}

\textbf{Concrete Example:}
We illustrate the result of our approach with Listing~\ref{code_instrumentation}.
It shows the transformation that the \textsc{Jimple} code undergoes (shown as Java code for readability).
The AICC method (program point of interest) appears at line 6.
After inferring the target component type with the help of \texttt{COAL}/\texttt{IC3}, \raicc{} generates a new standard ICC method call right after the AICC method (i.e., at line 7) corresponding to this type, i.e., \texttt{startActivity()}.
Indeed, the \texttt{PendingIntent} has been generated with the method \texttt{getActivity}, thus the target component type in the inferred values is defined as "a" in \texttt{COAL}/\texttt{IC3}, i.e., \texttt{Activity}.
Also, \raicc{} is able to recover the \texttt{Intent} used to create the \texttt{PendingIntent} for using it as a parameter for the new standard ICC method call.

\begin{listing}[h]
\begin{minted}{java}
  Intent  i = new Intent(this, TargetActivity.class);
  PendingIntent p = PendingIntent.getActivity(this, 0, i, 0);
  LocationManager l = (LocationManager)getSystemService("location");
  Criteria c = new Criteria();
  // program point of interest
  l.requestSingleUpdate(l.getBestProvider(c,false), p);
+ startActivity(i);
\end{minted}
    \caption{How \raicc{} would instrument an app. (Lines with "+" represent added lines)}
    \label{code_instrumentation}
\end{listing}

\textbf{Details of each step involved in \raicc:}\newline
\textbf{\emph{Step 1:}} The app is transformed into Jimple~\cite{vallee1998jimple}, the internal representation of the Soot framework~\cite{vallee2010soot} using Dexpler~\cite{bartel2012dexpler}.%, which is more suitable for generating an inter-procedural control-flow graph and facilitates static analysis.

\noindent
\textbf{\emph{Step 2:}} \raicc{} leverages IC3~\cite{octeau2015composite} %\jk{should we said that it is IC3 is an improved version of EPICC?} 
which is able to infer all possible values of ICC objects using composite constant propagation at specific program points.
To this end, we created model files using the COAL~\cite{octeau2015composite} declarative language to query each of the AICC methods during program analysis and retrieve the values of the parameters we need (i.e., \texttt{PendingIntent} and \texttt{IntentSender}).

Given that they are built from \texttt{Intent} objects, IC3 is able to identify all subparts which compose the objects (e.g., action, category, extras, URI, etc.).
The most important artifact for our instrumentation is the types of potential target components.
It is inferred by \texttt{COAL} given its specification, i.e., it is able to get the target component type by recognizing methods for creating \texttt{PendingIntents} (e.g., \texttt{getActivity}).
Indeed, one can easily see the difference between a conventional ICC method and an AICC method: standard ICC methods explicitly describe the type of component that will be launched (e.g., \texttt{startActivity()} for an activity, \texttt{startService()} for a service, \texttt{send\-Broadcast()} for BroadcastReceiver, etc.), whereas with AICC method we cannot statically directly know the type of those components (e.g., the signature of the \texttt{set()} method gives no information about the type of the target component, and it is the same for most of the AICC methods such as \texttt{sendTextMessage()}, \texttt{request\-LocationUpdates()}, etc.).

Depending on the control-flow of the program during execution, the target component can change, hence its type too.
Consequently, we have to take into account all possible types for different components.
The main idea of our instrumentation approach is to add as many new standard ICC method calls as there are target components types and \texttt{Intent} objects for creating \texttt{PendingIntent} and \texttt{IntentSender} right after the program points of interest.
The type is represented by a single character in the COAL specification for a given class.
For example, the target type of a \texttt{Pending\-Intent} can take the following values: 1) "a" for an \texttt{Activity}, 2) "r" for a \texttt{BroadcastReceiver} and 3) "s" for a \texttt{Service}.

\noindent
\textbf{\emph{Step 3:}}
After retrieving the possible target component types of the AICC methods, \raicc{} has to recover the right \texttt{Intent} that has been used for creating the \texttt{PendingIntent} or the \texttt{IntentSender} which will be the parameter of the generated standard ICC method(s).
To tackle this issue, \raicc{} first recovers the \texttt{PendingIntent} or \texttt{IntentSender} reference used in the AICC method.
Note that it can be used as a parameter in the AICC method (e.g., \texttt{sendText\-Message()}) or as the caller object (e.g., \texttt{send()}), we annotated each AICC method for having this information and, in the case it is a parameter, the index in the list of parameters.
Afterwards, \raicc{} interprocedurally searches for the \texttt{Intent} used for creating the \texttt{PendingIntent}.
In the case of \texttt{IntentSender}, \raicc{} interprocedurally searches for the \texttt{PendingIntent}, then recursively apply the previous process for retrieving the \texttt{Intent}.
Of course, different \texttt{Intent} objects could be used in the code (not shown in Listing~\ref{code_instrumentation}). Therefore for correctly propagating the "context information" among components for further analysis, they should all be taken into account, as \raicc{} does.

\noindent
\textbf{\emph{Step 4:}}
At this point, for each point of interest (AICC method), \raicc{} leverages the list of potential target component type and the list of potential \texttt{Intents}.
The source code modification of the app to explicitly set the ICC methods is straightforward.
After each AICC method, \raicc{} generates as many invoke statements as there are combinations of potential target types and potential \texttt{Intents} recovered.
The new generated invoke statements  will depend on the type(s) inferred at \emph{step 2}, i.e., \texttt{startActivity} for "a", \texttt{startService} for "s" and \texttt{sendBroadcast} for "r". 
\texttt{Intent} objects are used as parameters of the new method calls.

Note that some of the AICC methods, likewise \texttt{startActivity\-For\-Result()}, expect a result returned if the target component type is an \texttt{Activity}.
We have carefully annotated the corresponding AICC methods, therefore \raicc{} generates the right method call in this case, i.e., \texttt{startActivity\-ForResult()}.

\noindent
\textbf{\emph{Step 5:}}
Finally, \raicc{} packages the newly generated application, and any existing tool dealing with standard ICC methods can be used to perform further static analysis. % taint analysis to detect ICC data leaks.

Note that although instrumentation can lead to non-runnable apps, in this study, apps are not meant to be executed after being processed by \raicc{}. Indeed, \raicc{} acts as a preprocessor for other static analyses.

%% file: tikz/overview.tikz
\tikzset{node/.style={minimum height=0.6cm, rounded corners,scale=0.8, text width=2.3cm, minimum width=2cm,minimum height=1cm,align=center}}
\begin{tikzpicture}[scale=0.9]

\node[inner sep=0pt] (apk) at (-4.0,0) {\includegraphics[width=18pt]{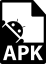}}; 
\node[node,draw] (jimple) at (-0.75,0) {1) Transform into Jimple};
\node[node,draw] (iiccm) at (2.5,0) {\small{2) Infer target component types}};
\node[node,draw] (target) at (2.5,-1.25) {3) Collect \texttt{Intents}};
\node[node,draw] (new_apk) at (-4.0,-1.25) {5) New APK generated};
\node[node,draw] (instru) at (-0.75,-1.25) {4) Code instrumentation};

\draw[->,>=latex] (apk) -- (jimple);
\draw[->,>=latex] (jimple) -- (iiccm);
\draw[->,>=latex] (iiccm) -- (target);
\draw[->,>=latex] (target) -- (instru);
\draw[->,>=latex] (instru) -- (new_apk);

\end{tikzpicture}

%% file: sections/evaluation.tex
\section{Evaluation}
\label{evaluation}
We address the following research questions:

\begin{description}
    \item[RQ1:] Do AICC methods deserve attention? In other words, are AICC methods often used in Android apps?
    \item[RQ2:] Are AICC methods new in the Android Ecosystem?
    \item[RQ3:] Can \raicc{} boost the precision of ICC-based data leak detectors on benchmark apps?
    \item[RQ4:] Does \raicc{} reveal previously undetected ICC links in real-world apps? If so, are these newly detected ICC links security-sensitive?
    \item[RQ5:] What are the runtime performance and the overhead introduced by \raicc{}?
\end{description}

\subsection{Atypical ICC Methods Deserve Attention}
\label{approach_indirect_icc_deserve_attention}

In section~\ref{approach_list_indirect_icc_methods}, we described how we build a list of \emph{atypical ICC methods}.
We used this list to conduct empirical analyses assessing the use of AICC methods in the wild.

In a first study, we randomly  selected 
\num{50000} malicious apps and \num{50000} benign apps from the Androzoo dataset~\cite{allix2016androzoo}. 
For qualifying the maliciousness of the apps, we used the VirusTotal~\cite{total2012virustotal} score 
(number of antivirus products that flag an app as malicious) available in the metadata of the app in Androzoo.
Every app of our malicious set has a VirusTotal score strictly greater than 20, those from the benign have a score equal to 0.

\textbf{Library code vs. developer code:} 
It has been shown ~\cite{li2016-libraries} than libraries present in Android apps can seriously impact empirical investigation performed on Android apps. 
Indeed, code related to libraries is often larger than the code written by the developers of the apps. 
For this reason, in this study, we perform two experiments: 
(1) we count the number of AICC methods present in each collected app by considering the entire code (i.e., including library code);
(2) we count the number of AICC methods only present in the developer code.
    In practice, to exclude library code, we rely on \texttt{Soot} which can discard third party libraries from a given list (in our experiments, we use the list from~\cite{li2016-libraries}) and system classes with simple heuristics (e.g., discard if the signature starts with "androidx.*" or "org.w3c.dom.*", etc.)

Table~\ref{empirical_AICCM} shows our findings.
We can see that among the benign apps, considering only the developer code, \num{24884} apps ($\sim$50\%) use at least one AICC method, and overall, \num{124226} AICC methods are used. 
    If we take into account the libraries, it is no less than \num{43754} apps (\num{87.5}\%) using in total \num{1154425} AICC methods.
    Clearly, in benign apps, the large majority of AICC methods are leveraged by libraries. 
In the malicious set, we face a different situation. 
The reported figures considering libraries or not are much closer. 
Finally, if we compare both datasets, we note that overall, benign apps tend to use much more AICC methods than malicious apps, 
but when considering only the code written by the developers of the apps, the situation is reversed, i.e., 
developers use much more AICC methods in malicious apps than in benign apps.

\begin{table}[h]
    \centering
\begin{adjustbox}{width=\columnwidth,center}
    \begin{tabular}{g|chc|hch}
        \hline
        \rowcolor{gray!70}
        & \multicolumn{3}{|c|}{\textbf{Without libs}} & \multicolumn{3}{|c}{\textbf{With libs}} \\
        \rowcolor{gray!50}
        \multicolumn{1}{l|}{\textbf{Dataset}} & \multicolumn{1}{|c}{\textbf{\# AICCM}} & \multicolumn{1}{c}{\textbf{\# apps}} & \multicolumn{1}{c|}{\textbf{ratio}$^\dagger$} & \multicolumn{1}{|c}{\textbf{\# AICCM}} & \multicolumn{1}{c}{\textbf{\# apps}} & \multicolumn{1}{c}{\textbf{ratio}$^\dagger$} \\ \hline \hline
        \multicolumn{1}{p{11mm}|}{\cellcolor{gray!30}50k \newline benign} & \multirow{2}{*}{\num{124226}} & \multicolumn{1}{p{13mm}}{\cellcolor{gray!30} \centering \num{24884} \newline (\num{49.8}\%)} &\multirow{2}{*}{\num{5}/app} & \multirow{2}{*}{\num{1154425}} & \multicolumn{1}{p{13mm}}{\centering \num{43754} (\num{87.5}\%)} & \multirow{2}{*}{\num{26.4}/app} \\
        \hline
        \multicolumn{1}{p{11mm}|}{\cellcolor{gray!30}50k \newline malicious } & \multirow{2}{*}{\num{402468}} & \multicolumn{1}{p{13mm}}{\cellcolor{gray!30} \centering \num{34710} \newline (\num{69.4}\%)} &\multirow{2}{*}{\num{11.6}/app} & \multirow{2}{*}{\num{522126} }& \multicolumn{1}{p{13mm}}{\centering \num{39845} (\num{79.7}\%)} & \multirow{2}{*}{\num{13.1}/app} \\
        \hline
    \end{tabular}
\end{adjustbox}
{$^\dagger$\footnotesize The ratio is computed by considering apps with at least one AICC method.}
        \caption{Number of apps using at least one AICC method in different datasets (AICCM: AICC method).}
    \label{empirical_AICCM}
\end{table}

\vspace{-3mm}
Table~\ref{table_most_used_aiccm} presents for both datasets the top 5 used AICC methods in developer code (excluding libraries). 
We notice 3 common AICC methods in this table (i.e., \texttt{set}, \texttt{setRepeating} and \texttt{setLatest{\-}EventInfo}).
Regarding the malicious apps, we can see that the methods from the class \texttt{SmsManager} are present twice. % in the top 5.
It could be explained by the fact that malicious apps tend to activate components via SMS. % and/or send premium SMS.
We also note that method \texttt{set{\-}Latest{\-}Event{\-}Info} is used an order of magnitude more than all other methods. 
This method is actually related to the notification mechanism of Android. 
We postulate that malicious apps tend to be much more aggressive in terms of notifications and advertisements resulting in a high number of usages of this method. 

\begin{table}[h]
    \begin{adjustbox}{width=0.95\columnwidth,center}
    \begin{tabular}{lcc}
        \hline
        \rowcolor{gray!70}
        \multicolumn{1}{l}{\textbf{Methods}} & \multicolumn{1}{c}{\textbf{Counts}} & \multicolumn{1}{c}{\textbf{\%}} \\ \hline \hline
        \rowcolor{gray!40}
        \multicolumn{3}{c}{\textbf{Benigns} (\num{50000})} \\ \hline
        android.app.AlarmManager.set& \num{27214} & \num{21.9}\% \\ \hline
        android.widget.RemoteViews.setOnClickPendingIntent & \num{19217} & \num{15.5}\% \\ \hline
        android.app.Notification.setLatestEventInfo & \num{18024} & \num{14.5}\% \\ \hline
        android.app.AlarmManager.setRepeating & \num{9184} & \num{7.4}\% \\ \hline
        android.app.Activity.startIntentSenderForResult & \num{6876} & \num{5.5}\% \\ \hline
        %android.support.v4.app.NotificationCompat\$Builder.addAction & \num{5357} & \num{}\% \\ \hline
        %android.app.AlarmManager.setInexactRepeating & \num{4144} & \num{}\% \\ \hline
        %android.app.PendingIntent.send & \num{4010} & \num{}\% \\ \hline
        %android.app.Notification\$Builder.setDeleteIntent & \num{3819} & \num{}\% \\ \hline
        %android.app.AlarmManager.setExact & \num{3723} & \num{}\% \\ \hline
        \rowcolor{gray!40}
        \multicolumn{3}{c}{\textbf{Malicious} (\num{50000})} \\ \hline
        android.app.Notification.setLatestEventInfo & \num{238462} & \num{59.2}\% \\ \hline
        android.app.AlarmManager.set & \num{53533} & \num{13.3}\% \\ \hline
        android.telephony.SmsManager.sendTextMessage & \num{39011} & \num{9.7}\% \\ \hline
        android.app.AlarmManager.setRepeating & \num{22813} & \num{5.7}\% \\ \hline
        android.telephony.SmsManager.sendDataMessage & \num{13075} & \num{3.2}\% \\ \hline
    \end{tabular}
    \end{adjustbox}
    \caption{Most used atypical ICC methods in benign/malicious Android apps, without considering libraries.}
    \label{table_most_used_aiccm}
\end{table}

\begin{figure}[h]
\vspace{-0.3cm}
\centering
       \includegraphics[scale=.56]{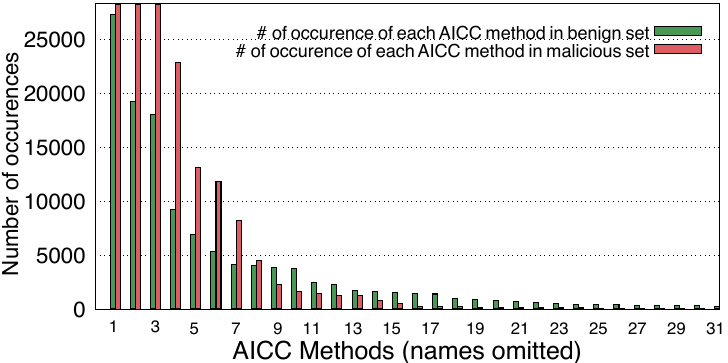}
\caption{Occurence of AICC methods in benign and malicious applications (excluding libraries)}
\label{fig:plot_aiccms_usage}
\vspace{-0.3cm}
\end{figure}

Finally, Figure~\ref{fig:plot_aiccms_usage} presents the number of usages of each of the 111 AICC methods in the developer code in both benign and malicious datasets.
For each dataset, the methods are ranked by their number of occurrences. 
For the sake of readability, we have truncated the first two bars of the malicious datasets. 
Indeed, as shown in Table~\ref{table_most_used_aiccm}, the number of occurrence of the top 3 methods are $\sim$238k, $\sim$53k and $\sim$39k respectively.
Thanks to Figure~\ref{fig:plot_aiccms_usage}, we note that: 
(1) only a fraction of the AICC methods is largely used by developers,
(2) 21 methods are even not used at all,
(3) malicious developers tend to use a less diverse set of AICC methods but the AICC methods that are used, are more frequently used.

\highlight{\textbf{RQ1 Answer:} 
AICC methods are prevalent in Android apps, and thus definitely deserve attention. 
They are used in both malicious and benign apps, but significantly more by malicious developers.
Only a fraction of the AICC methods are regularly used. 
}
\vspace{-2mm}

\subsection{Atypical ICC Methods exist since the beginning}
\label{subsec:evolution}
\vspace{-1mm}
To the best of our knowledge, state of the art approaches do not consider AICC methods. 
One of the reasons could be the fact that AICC methods have only been introduced recently in the Android Framework. 
To validate this hypothesis, we further check the use of AICC methods over time. 
For this purpose, we considered 5 sets of \num{5000} benign apps from Androzoo (ordered by the creation date of the dex file), %on which we run the same analysis.
and 4 sets of malicious apps.
Androzoo only contains a few malicious apps from 2019 and no malicious app from 2020.
Thus, the 2019 malicious set is reduced compared to the benign one and there is no 2020 malicious set.
The sets, their content and the results of the analyses are provided in Table~\ref{empirical_AICCM_temporal}.

First, overall these results confirm the results of Table~\ref{empirical_AICCM}. 
For instance, in benign apps, AICC methods are mostly used in libraries.
Malicious developers still use more AICC methods in their code, even if the difference between with or without libraries is less pronounced.
Regarding temporal evolution, we note that in both datasets, the metrics are pretty stable,  
except maybe in 2019 -malicious set- which seems to be an outlier (weak ratio and high \% of number of apps). This could be explained by the low number of apps (548) collected for 2019. 

\begin{table}[h]
    \centering
\begin{adjustbox}{width=\columnwidth,center}
    \begin{tabular}{gchchch}
        \hline
        \rowcolor{gray!70}
        & \multicolumn{3}{c|}{\textbf{Withtout libs}} & \multicolumn{3}{|c}{\textbf{With libs}} \\
        \rowcolor{gray!70}
        \multicolumn{1}{l}{\textbf{Dataset}} & \multicolumn{1}{c}{\textbf{\# AICCM}} & \multicolumn{1}{c}{\textbf{\# apps}} & \multicolumn{1}{c|}{\textbf{ratio}$^\dagger$} & \multicolumn{1}{|c}{\textbf{\# AICCM}} & \multicolumn{1}{c}{\textbf{\# apps}} & \multicolumn{1}{c}{\textbf{ratio}$^\dagger$} \\ \hline \hline
        \rowcolor{gray!40}
        \multicolumn{7}{c}{\textbf{Benign Sets}} \\
        2016 (5000) & \num{14130} & \num{2620} (\num{52.40}\%) & \num{5.4}/app & \num{129089} & \num{4584} (\num{91.68}\%) & \num{28.2}/app \\
        2017 (5000) & \num{11540} & \num{2486} (\num{49.72}\%) & \num{4.6}/app & \num{133803} & \num{4601} (\num{92.02}\%) & \num{29.1}/app \\
        2018 (5000) & \num{15167} & \num{2487} (\num{49.74}\%) & \num{6.1}/app & \num{143009} & \num{4708} (\num{94.16}\%) & \num{30.4}/app \\
        2019 (5000) & \num{15923} & \num{2629} (\num{52.58}\%) & \num{6.0}/app & \num{144467} & \num{4528} (\num{90.56}\%) & \num{31.9}/app \\
        2020 (5000) & \num{15300} & \num{2403} (\num{48.06}\%) & \num{6.4}/app & \num{106019} & \num{3488} (\num{69.76}\%) & \num{30.4}/app \\
        \rowcolor{gray!40}
        \multicolumn{7}{c}{\textbf{Malicious Sets}} \\
        2016 (5000) & \num{20156} & \num{2371} (\num{47.42}\%) & \num{8.5}/app & \num{58967} & \num{2997} (\num{59.94}\%) & \num{19.7}/app \\
        2017 (2825) & \num{16316} & \num{1222} (\num{43.26}\%) & \num{13.3}/app & \num{45832} & \num{1583} (\num{56.03}\%) & \num{28.9}/app \\
        2018 (3067) & \num{28083} & \num{1676} (\num{54.65}\%) & \num{16.8}/app & \num{56623} & \num{1823} (\num{59.44}\%) & \num{31.1}/app \\
        2019 (548) & \num{1494} & \num{378} (\num{68.98}\%) & \num{3.9}/app & \num{7268} & \num{429} (\num{78.28}\%) & \num{16.9}/app \\
        \hline
    \end{tabular}
\end{adjustbox}
{$^\dagger$\footnotesize The ratio is computed by considering apps with at least one AICC method.}
    \caption{Temporal evolution of the usage of AICC methods in benign and malicious apps. }
    \label{empirical_AICCM_temporal}
\end{table}

\begin{figure}[h]
\centering
       \includegraphics[scale=0.50]{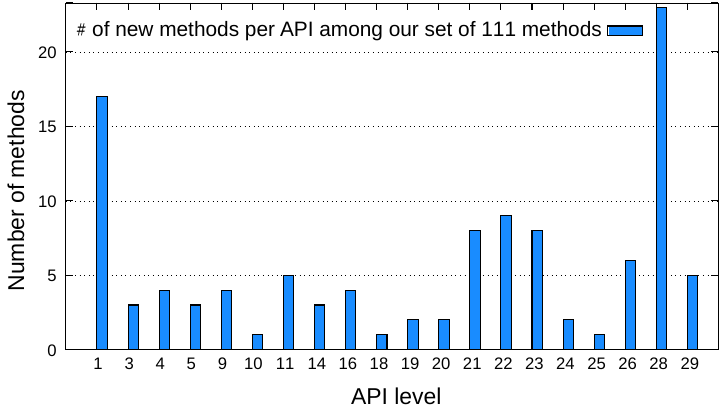}
       \caption{API levels in which AICC methods have been added.}
    \label{plot_apis}
\end{figure}

To deepen our investigation about temporal evolution, we also study the "introduction time" of the 111 AICC methods. 
To that end, we count the number of AICC methods introduced in each Android API level. 
The results are presented in Figure~\ref{plot_apis}. 
New AICC methods have been added at almost each API level (often between 1 and 5 per API level). 
We can see two peaks: one at API level 1 corresponding to the creation of the Android Framework, 
and one at API level 28 corresponding to the introduction of AndroidX, a new set of Android libraries. 
It is noteworthy that only two AICC methods have been removed from the Android framework.

\vspace{-2mm}

\highlight{\textbf{RQ2 Answer:} 
AICC methods are not new in the Android framework, they indeed exist since the very beginning. 
}
\vspace{-1mm}

\subsection{Precision improvement  after applying \raicc{}}
\label{subsec:benchmark}

RQ3 aims at investigating the efficiency of state-of-the-art ICC data leak detector \iccta{} and \textsc{Amandroid} after applying \raicc{}.
To do so, we launched the tools before and after executing \raicc{} against 20 new apps that we plan to integrate into \droidbench{}~\cite{arzt2014flowdroid},
an open test suite containing more than 200 hand-crafted Android apps for evaluating the efficiency of taint analyzers.
\droidbench{} is used as a ground-truth by the research community in order to assess the efficiency of static and dynamic analyzers.
It contains different types of leaks, e.g., intra-component, inter-component, inter-app, etc.
However, among the ICC leaks, none of them uses AICC methods.
Thus, our idea is to extend \droidbench{} with 20 additional test cases focusing on ICC leaks (concrete application of taint tracking) performed via AICC methods.
Note that, to detect false positives, we included 4 apps without leak among the 20 apps (i.e., only 16 apps contain a leak).

\noindent
\textbf{Benchmark construction}
To develop those 20 apps, we considered the most representative AICC methods for both malicious and benign apps identified in Section~\ref{approach_indirect_icc_deserve_attention}.
More specifically, we considered the top 10 AICC methods (in terms of occurrences) in both datasets leading to 14 AICC methods (10+10-6 duplicates). 
We also randomly picked 4 additional AICC methods to reach the final number of 18 AICC methods (2 AICC methods have been used twice), which represent 
 93.5\%  and  91.1\%  of the AICC methods occurrences in our datasets of \num{50000} benign apps and \num{50000} malicious apps respectively. 

The implementation of most of our bench apps was straightforward as well as the triggering of the underlying inter-component communication. 
Excerpts of such bench apps are similar to the ones presented in Listing~\ref{code_iiccm_examples}.
However, some AICC methods have required more sophisticated code, e.g., those manipulating \texttt{Notification} objects, for instance the \texttt{addAction} AICC method.
Another example of more complex bench app is related to the AICC method  \texttt{setOnClickPendingIntent} of the \texttt{android.widget.\-RemoteViews} class.
The \texttt{PendingIntent} set as parameter of this method is triggered after the user clicks on a widget appearing in the home screen of the device.
The widget (declared in the AndroidManifest.xml file) has to be installed on the home screen before the user can click on it to trigger the target component.

Note that, beside developing applications using AICC methods, we combine multiple aspects of the way ICC can be performed.
For example, in several apps, we considered the data flow within three different components or a data flow looping back into the first component to check the behavior or \raicc{}.

Table~\ref{droidbench_descripton} lists the 20 bench apps. 
For the sake of space, we cannot give more details about this benchmark, but we invite the interested reader to refer to the project repository\footnote{https://github.com/Trustworthy-Software/RAICC} which contains the source code of each bench app.

\begin{table}[h]
{\small$\ostar$ = true-positive, \textcolor{red}{$\star$} = false-positive, \textcolor{red}{$\bigcirc$} = false-negative,  \textbf{C} = Components, UI = User Interaction} \\
 \begin{adjustbox}{width=\columnwidth,center}
\begin{tabular}{lccccccc}
    \hline
        \rowcolor{gray!70}
    \bf Test Case & \bf \# C. & \bf Leak & \bf UI & \multicolumn{2}{|c|}{\textbf{\iccta{} }}  & \multicolumn{2}{|c|}{\textbf{\footnotesize Amandroid}} \\
    \hline
    \hline
    \rowcolor{gray!30}
sendTextMessage1  & 2 & $\bullet$  & $\circ$ & \textcolor{red}{$\bigcirc$} & $\ostar$ & \textcolor{red}{$\bigcirc$} & $\ostar$ \\
setSendDataMessage & 3 & $\bullet$ &  $\circ$ & \textcolor{red}{$\bigcirc$} & $\ostar$ & \textcolor{red}{$\bigcirc$} & $\ostar$ \\
\rowcolor{gray!30}
sendTextMessage2  & 2  & $\circ$  & $\circ$ & & \textcolor{red}{$\star$}  & & \textcolor{red}{$\star$} \\
addAction1  & 2   & $\bullet$  & $\bullet$ & \textcolor{red}{$\bigcirc$} & $\ostar$ & \textcolor{red}{$\bigcirc$} & $\ostar$ \\
\rowcolor{gray!30}
addAction2  & 2   & $\circ$  & $\bullet$ & & \textcolor{red}{$\star$} & & \textcolor{red}{$\star$} \\
requestNetwork & 2   & $\bullet$  & $\circ$ & \textcolor{red}{$\bigcirc$} & $\ostar$ & \textcolor{red}{$\bigcirc$} & $\ostar$ \\
\rowcolor{gray!30}
requestLocationUpdates & 3   & $\bullet$  & $\circ$ & \textcolor{red}{$\bigcirc$} & $\ostar$ & \textcolor{red}{$\bigcirc$} & $\ostar$ \\
startIntentSenderForResult  & 2  & $\bullet$ & $\circ$ & \textcolor{red}{$\bigcirc$} & $\ostar$ & \textcolor{red}{$\bigcirc$} & \textcolor{red}{$\bigcirc$} \\
\rowcolor{gray!30}
send   & 3   & $\circ$  & $\circ$ &  & & & \textcolor{red}{$\star$} \\
sendIntent  & 2   & $\circ$ & $\circ$ &                            &  & &   \\
\rowcolor{gray!30}
setRepeating  & 2   & $\bullet$ & $\circ$ & \textcolor{red}{$\bigcirc$} & $\ostar$ & \textcolor{red}{$\bigcirc$} & $\ostar$ \\
setOnClickPendingIntent & 3  & $\bullet$ & $\bullet$  & \textcolor{red}{$\bigcirc$} & $\ostar$ & \textcolor{red}{$\bigcirc$} & $\ostar$ \\
\rowcolor{gray!30}
setLatestEventInfo  & 2   & $\bullet$ & $\bullet$ & \textcolor{red}{$\bigcirc$} & $\ostar$ & \textcolor{red}{$\bigcirc$} & $\ostar$\\
setInexactRepeating & 2   & $\bullet$ & $\circ$ & \textcolor{red}{$\bigcirc$} & $\ostar$ & \textcolor{red}{$\bigcirc$} & $\ostar$ \\
\rowcolor{gray!30}
setExact  & 2   & $\bullet$  & $\circ$ & \textcolor{red}{$\bigcirc$} & $\ostar$ & \textcolor{red}{$\bigcirc$} & $\ostar$  \\
setExactAndAllowWhileIdle & 2  & $\bullet$ & $\circ$ & \textcolor{red}{$\bigcirc$} & $\ostar$ & \textcolor{red}{$\bigcirc$} & $\ostar$  \\
\rowcolor{gray!30}
setWindow & 2   & $\bullet$ & $\circ$ & \textcolor{red}{$\bigcirc$} & $\ostar$ & \textcolor{red}{$\bigcirc$} & $\ostar$ \\
setDeleteIntent & 2   & $\bullet$  & $\bullet$ & \textcolor{red}{$\bigcirc$} & $\ostar$ & \textcolor{red}{$\bigcirc$} & $\ostar$ \\
\rowcolor{gray!30}
setFullScreenIntent & 2   & $\bullet$ & $\bullet$ & \textcolor{red}{$\bigcirc$} & $\ostar$ & \textcolor{red}{$\bigcirc$} & $\ostar$ \\
setPendingIntentTemplate & 3   & $\bullet$ & $\bullet$ & \textcolor{red}{$\bigcirc$} & $\ostar$ & \textcolor{red}{$\bigcirc$} & $\ostar$ \\
\rowcolor{gray!50}
\multicolumn{8}{c}{Sum, Precision, Recall} \\
$\ostar$, higher is better & & & & 0 & 16 & 0 & 15 \\
\rowcolor{gray!30}
\textcolor{red}{$\star$}, lower is better & & &  & 0 & 2 & 0 & 3 \\
\textcolor{red}{$\bigcirc$}, lower is better & & &  & 16 & 0 & 16 & 1 \\
\rowcolor{gray!30}
Precision $p = \ostar / (\ostar + $ \textcolor{red}{$\star$} $)$ & & &  & 0\% & 88.90\% & 0\% & 83.33\% \\
Recall $r = \ostar / ( \ostar + $ \textcolor{red}{$\bigcirc$} $)$ & & &  & 0\% & 100\% & 0\% & 93.75\% \\
\rowcolor{gray!30}
$F_1$-score $= 2pr/(p+r)$ & & &  & 0 & 0.94 & 0 & 0.88 \\
    \hline
\end{tabular}
\end{adjustbox}
    \caption{Additional \droidbench{} apps and results of applying \iccta{} and Amandroid before and after \raicc{}.
    \footnotesize A more complete Table is available in the supplementary material document.}
    \label{droidbench_descripton}
\end{table}

\noindent
\textbf{Results}
Table~\ref{droidbench_descripton} shows the results of our experiment.
Since \iccta{} and \textsc{Amandroid} are not designed to detect ICC data leaks via AICC methods, 
it is not surprising to see that they performs very badly without applying \raicc{} (precision and recall of 0\%).
Indeed, \iccta{} and \textsc{Amandroid} are not able to construct the links between the components for the 16 apps containing a leak.
However, for the 4 apps which do not contain any leak, they do not raise any alarm as expected.

After instrumenting the apps with \raicc{}, the performance of \iccta{} and \textsc{Amandroid} is improved.
They can reveal and construct previously hidden ICC enabling the detection of the leaks present in this benchmark.

Regarding \iccta{}, it is able to reveal all the leaks after applying \raicc{}.
However, we can see 2 false-positives.
The first one, in app "sendTextMessage2", is due to \iccta{} which cannot correctly parse extra keys added into \texttt{Intent} objects (cf. \texttt{startActivity7 of \droidbench}).
The second one is due to \raicc{} which cannot, for the moment, differentiate atypical inter-com\-ponent communication made asynchronously.
What we mean is that in "addAction2", the notification is never shown to the user, hence the component targeted by the \texttt{PendingIntent} will not be executed through the notification.
Therefore, the leak cannot happen during execution.
Even if declaring a notification and not showing it to the user is not likely to happen in practice, it is a good example to show that modeling an app behavior is not trivial and demands more effort for certain methods.
We can notice that \iccta{} behaves correctly with apps "send" and "sendIntent" by not raising an alarm.

\textsc{Amandroid} performance is also boosted. Indeed, it can reveal almost all the leaks (1 false-negative).
We can notice that the same false-positives appears for \iccta{} and \textsc{Amandroid} for apps "sendTextMessage2" and "addAction2".
\textsc{Amandroid} reveals an additional false-positive for app "send".

As a result, the precision of \iccta{} combined with \raicc{} reaches 88.90\% (16 true-positives and 2 false-positives) and its recall 100\% (16 true-positives and 0 false-negative).
As for \textsc{Amandroid}, combined with \raicc{} its precision reaches 83.33\% (15 true-positives and 3 false-positives) and its recall 93.75\% (15 true-positives and 1 false-negative).
\iccta{} $F_1$-score reaches 0.94 and \textsc{Amandroid} 0.88.

\highlight{
    \textbf{RQ3 Answer:} \raicc{} boosts both the precision and the recall of state-of-the-art data leak detectors.}% \iccta{} and \texttt{Amandroid}.

\subsection{Experimental results on real-world apps}

In this section, we first investigate to what extent \raicc discovers previously undetected ICC links in real-world apps. 
Then, we perform two checks on these newly detected ICC links: 
(1) we check if they are used to transfer data across components or even to perform some privacy leaks;
(2) we check if they lead to ICC vulnerabilities.

\subsubsection{Revealing new ICC links}
In this section, we study the capacity of \raicc{} in revealing new ICC links in real-world apps. 
To that end, we extract, from Androzoo, two datasets of  \num{5000} randomly selected apps containing respectively only benign and malicious apps. 
Then for each app, we count the number of ICC links discovered without \raicc{} (by relying on the results yielded by \ic{}), 
as well as the number of additional ICC links discovered by \raicc{}.
Note that we only consider the developer code in this study (i.e., we exclude the libraries).
Table~\ref{table_type_components} presents our results.

\begin{table}[h]
    \centering
 \resizebox{0.88\linewidth}{!}{    
    \begin{tabular}{g|ch|chr}
        \hline
        \rowcolor{gray!70}
         & \multicolumn{2}{|c|}{\textbf{\ic{} }}  & \multicolumn{2}{|c|}{\textbf{\raicc{} }} &\multicolumn{1}{|c}{\textbf{Increase}} \\ 
        \rowcolor{gray!70}
        \multicolumn{1}{l|}{\textbf{Component types}} & \multicolumn{1}{|c}{\textbf{Counts}} & \multicolumn{1}{c|}{\textbf{\%}} & \multicolumn{1}{|c}{\textbf{Counts}} & \multicolumn{1}{c|}{\textbf{\%}} &\multicolumn{1}{|c}{\textbf{\%}}  \\ \hline \hline
        \rowcolor{gray!50}
        \multicolumn{6}{c}{Benign set (\num{5000})} \\
        Activity & \num{17095} & \num{84.2}\% & +\num{2463} & \num{45.5}\% & +\num{14.4}\% \\ 
        BroadcastReceiver& \num{1221} & \num{6.0}\% & +\num{1907} & \num{35.3}\% & +\num{156.2}\% \\ 
        Service & \num{1984} & \num{9.8}\% & +\num{1038} & \num{19.2}\% & +\num{52.3}\% \\ 
        Total & \num{20300} & \num{100}\% & +\num{5408} & \num{100}\% & +\num{26.6}\% \\ 
        \rowcolor{gray!50}
        \multicolumn{6}{c}{Malicious set (\num{5000})} \\ 
        Activity & \num{13489} & \num{83.1}\% & +\num{7340} & \num{73.4}\% & +\num{54.4}\% \\ 
        BroadcastReceiver & \num{747} & \num{4.6}\% & +\num{1468} & \num{14.7}\% & +\num{196.5}\% \\ 
        Service & \num{1986} & \num{12.3}\% & +\num{1193} & \num{11.9}\% & +\num{60.1}\% \\ 
        Total & \num{16222} & \num{100}\% & +\num{10001} & \num{100}\% & +\num{61.6}\% \\ 
    \end{tabular}
 }
    \caption{Number of ICC links resolved by \ic{} and number of additional ICC links discovered by  \raicc{}.}
    \label{table_type_components}
\end{table}

Among \num{5000} benign apps, \num{5408} new ICC links were revealed by \raicc{}, corresponding to an increase of more than 25\% in comparison with \ic.
The most used target component type is \texttt{Activity} with 45\% of the new links. 
However, while for \ic{} the large majority of ICC links are related to \texttt{Activity}, the distribution among the 3 types of component is more balanced with \raicc.
As regards to malicious apps, 
while the number if ICC links revealed by IC3 is relatively close to the number of ICC links revealed in the benign dataset (\num{20300} vs. \num{16222}), 
the number of ICC links revealed by \raicc{} is much higher, almost twice as much as benign apps (5408 vs. \num{10001}).
Overall, \raicc{} increases the number of ICC links by 61\% in malicious apps.

All three types of components are impacted by \raicc.   
However, the increase of the number of ICC links is impressive for \texttt{Broad{\-}castReceiver}: 156\% for benign apps, and almost 200\% for malicious apps. 
This suggests that developers tend to use AICC methods more than traditional ICC methods to "broadcast" an event. 
Through manual inspection, we indeed notice that, for instance, an AICC method attached to an "alarm" is often used to trigger a \texttt{Broad{\-}castReceiver}.

Finally, note that we also randomly picked 40 benign and malicious apps to manually verify if \raicc{} had correctly instrumented the real-world apps.
The standard ICC methods are correctly added right after the AICC methods, allowing other tools to correctly model ICC.

\subsubsection{Atypical ICC methods are largely used in real-world apps, although not to transfer or leak data}

For this study, we only consider a set of 5000 malicious apps 
(the underlying intuition is that malicious apps tend to leak more data than benign apps). 
We first run \raicc{} on this dataset (to resolve the atypical ICC links), and then we leverage \iccta{} to perform the detection of ICC leaks (\iccta{} uses a set of well-defined sources (i.e. sensitive information) and sinks to perform the detection). 
Overall, \iccta{} was able to detect 6129 intra-component data leaks (i.e., leaks inside a single method) and 114 ICC data leaks.
We manually inspect all 114 ICC data leaks to check if the data is transferred via AICC methods or standard ICC methods such as \texttt{startActivity()}. 
We did not find a single case where sensitive information is leaked via AICC methods.

We manually analyzed 60 apps to verify how AICC methods were used.
In the majority of cases the target component is used likewise a callback method, i.e., this mechanism is used to "activate" a given component.
Actually, when data is put inside the \texttt{Intent} used for constructing the \texttt{PendingIntent} or the \texttt{IntentSender}, it is generally non-sensitive data (most of the time simple  constants).
Let us consider a concrete example, for instance the "M1 Trafik" app
from the Google PlayStore.
In method \texttt{setAlarm} of class 
{\small \texttt{com.m1\_trafik.Alarm\-ManagerBroadcast\-Receiver}}, 
an \texttt{Intent} is created with an extra value representing the \texttt{Boolean} false value. 
Information attached to this intent also informs us that the target component is the current class itself (i.e. the class \texttt{Alarm\-ManagerBroad\-cast\-Receiver}).
A \texttt{PendingIntent} is then retrieved from this \texttt{In\-tent} using method \texttt{getBroadcast()}.
Afterwards, the AICC method \texttt{setRepeating()}  of class \texttt{AlarmManager} is leveraged for setting an alarm.
When this alarm goes off, the method \texttt{onReceive} of the target component (in our case the same class) is executed.
When analyzing this method we can see no use of the extra value put in the \texttt{Intent}.
When applying \raicc{}, we can see the new method call \texttt{sendBroadcast()} right after the call to \texttt{setRepeating()}.
Although it helps \iccta{} constructing the link between the components, the data transferred is not sensitive.
In this example, we see that AICC methods are mostly used to leverage the powerful "token" mechanism explained in Section~\ref{motivation}, i.e., the target component will be launched even if the application is closed.

\subsubsection{\raicc \& \textsc{EPICC} - revealing new ICC vulnerabilities}
\textsc{EPICC}~\cite{octeau2013effective} is a state-of-the-art ICC links resolver able to detect ICC vulnerabilities.
Such vulnerabilities are defined by Chin et al. in~\cite{chin2011analyzing}. 
Examples include (1) when an app sends an Intent that may be intercepted by a malicious component, or (2) when legitimate app components, --e.g., a component sending sms messages-- are activated via malicious \textit{Intent}. 
In this section we aim at showing that \raicc boosts \textsc{EPICC} by 
enabling the detection of previously unnoticed ICC vulnerabilities. 
To this end, we considered a dataset of 1000 randomly selected benign apps, 
and a dataset of 1000 randomly selected malicious apps.
We ran \textsc{EPICC} on those two datasets before and after applying \raicc, results are available in Table~\ref{table_vulnerabilities}.

\begin{table}[h]
    \begin{adjustbox}{width=0.80\columnwidth,center}
    \begin{tabular}{lcc}
        \hline
        \rowcolor{gray!70}
        & \multicolumn{1}{l}{\textbf{1000 benign apps}} & \multicolumn{1}{c}{\textbf{1000 malicious apps}}\\ \hline
        Before \raicc & \num{4796} & \num{9544} \\
        \rowcolor{gray!30}
        After \raicc & \num{5032} & \num{9868} \\
        Improvement & +236 (+4.9\%) & +324 (+3.4\%) \\
        \hline
    \end{tabular}
    \end{adjustbox}
    \caption{Number of ICC vulnerabilities found by \textsc{EPICC} before and after applying \raicc}
    \label{table_vulnerabilities}
    \vspace{-0.3cm}
\end{table}

Besides the significant difference between benign and malicious apps, % (almost a two-order-of-magnitude difference), 
we can see that after applying \raicc, i.e., modeling previously unrevealed ICC links, \textsc{EPICC} is able to detect more ICC vulnerabilities, 
with an increase of 4.9\% for benign apps and 3.4\% for malicious apps.
This experiment shows that \raicc boosts state-of-the-art tool \textsc{EPICC} by modeling new ICC links and revealing new ICC vulnerabilities.

\highlight{\textbf{RQ4 Answer:} \raicc{} significantly increases the number of resolved ICC links in real-world apps compared to the state-of-the-art approach.
While AICC methods seem to not be used to leak sensitive information, they are used to activate components (and thus potentially trigger malicious payloads).
\raicc boosts \textsc{EPICC} by allowing to reveal new ICC vulnerabilities.}

\subsection{Runtime performance of \raicc{}}
\label{eval_runtime}

In this section, we evaluate the runtime performance of \raicc{}. 
We also evaluate the overhead introduced by our tool by considering a typical usage of \raicc{}, for instance when \raicc{} is used to boost the results of \iccta.
Since \iccta{} leverages itself \ic{}, we investigate the runtime performance of   \ic{} and \iccta{}  before and after applying 
\raicc{} on the 10 benchmark apps used in Section~\ref{subsec:benchmark}.

\begin{figure}[h]
\centering
       \includegraphics[scale=0.34]{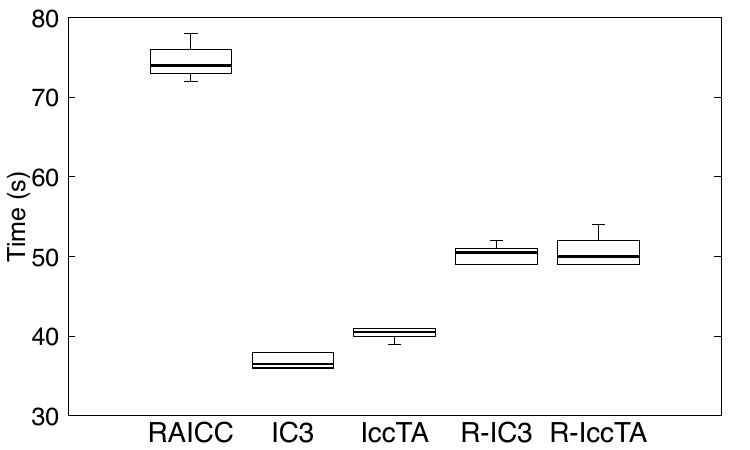}
       \includegraphics[scale=0.34]{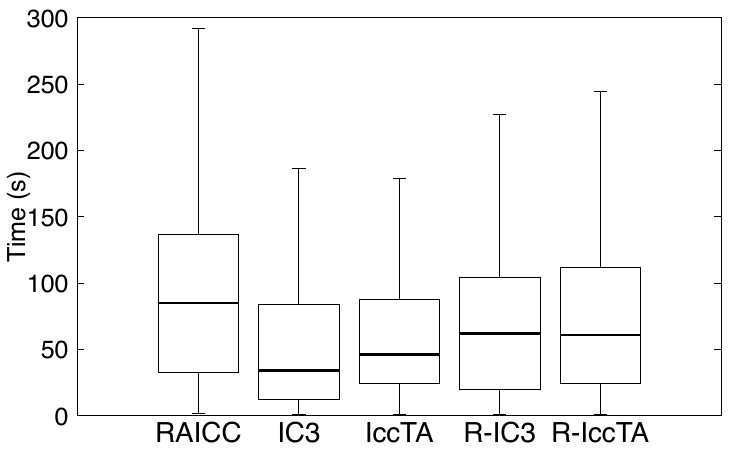}
       \caption{Runtime performance of RAICC, IC3 and IccTA with (R- means with \raicc) and without AICCM preprocessing. (left: on Droidbench, right: in the wild).}
    \label{fig:plot_runtime}
\end{figure}

The results are presented in Figure~\ref{fig:plot_runtime}. 
First, we can see that the \raicc{} execution time does not exceed 80 seconds. % for instrumenting an app.
Since \raicc{} allows \ic{} and \iccta{} to resolve more additional ICC links, we expect that the analysis time of both tools will increase. 
We indeed note that the two box-plots on the right are higher confirming the overhead caused by \raicc{}. 
On average, the overheads for \ic{} and \iccta{} are \num{13.3} seconds and \num{10} seconds  respectively (\num{36.74}\% and \num{24.74}\% overhead respectively).

To confirm the results obtained on the benchmark apps, we perform the same study but on a set of  1000 real-world apps.
The results are reported in Figure~\ref{fig:plot_runtime}.
Overall, we can see that the performances (in time) reported on both figures are quite similar. 
However, slight differences can be noticed. First, the runtime values are more scattered in Figure~\ref{fig:plot_runtime}  than with the benchmark apps. This could be explained by the fact that real-world apps are more diverse. 
Second, the average performances of the three tools are closer.

Regarding the overhead introduced by \raicc{}, we again notice that this overhead exists.
This is expected since the constant propagation of \ic{} has to process more values/methods.
Likewise, \iccta{} has to build more links and to consider more paths for the taint analysis.
On this dataset, on average, the overheads for \ic{} and \iccta{} are \num{21.8} seconds and \num{5.8} seconds respectively (+\num{43.8}\% and +\num{6.5}\% overhead respectively).

 \highlight{
 \textbf{RQ5 Answer:}   The runtime performance of \raicc{} is higher than \ic{} and \iccta{}, but still in the same order of magnitude. On average, \raicc{} requires less than 2 minutes to analyze and instrument a real-world application.
    }

%% file: sections/limitations.tex
\section{Limitations}
\label{limitations}

The core component of our approach lies in the list of AICC methods that we compiled during our research.
Even though we followed a systematic approach for retrieving a maximum of AICC methods, we might have missed some of them in the Android Framework.
There are potentially other ways to perform such ICC, nevertheless, our study is reproducible and provides insight for future research in this direction.

By leveraging IC3 to infer the values of ICC objects, RAICC inherits the limitations of IC3. 
Moreover, like most of the static analysis approaches, \raicc{} is subject to false-positives.
Currently, \raicc{} does not handle native calls, reflective calls nor dynamic class loading, though some state-of-the-art approach could be integrated~\cite{bodden2011taming, li2016droidra}.
Besides, although inter-app communication (IAC) is performed using the same mechanisms as 
ICC~\cite{li2015apkcombiner}, we did not investigate in this direction.

Furthermore, obfuscation is a confounding factor impacting studies based on APKs~\cite{10.1145/3180155.3180228, 10.1145/2597073.2597109}. Therefore, \raicc{}'s effectiveness is impacted by obfuscated code, especially if AICC method calls are disguised (e.g., using reflection).

%% file: sections/related_work.tex
\section{Related work}
\label{related_work}

To the best of our knowledge, we have presented the first approach taking into account AICC methods for connecting Android components. 
However, as explained in a systematic literature review~\cite{li-2017}, the research literature has proposed a large body of works focusing on statically analyzing Android apps. 
One of the most popular topics is the use of static analysis for checking security properties, and in particular for checking data leaks. 
The pioneer tools such as FlowDroid, Scandal, and others ~\cite{arzt2014flowdroid, gibler2012androidleaks, kim2012scandal, mann2012framework, yang2012leakminer} have started to focus on the detection of intra-component data leaks.
They all face the limitations of not being able to detect ICC leaks.

Several approaches have been developed to perform data leak detection between components. We will present these approaches in the following. 
\textbf{\textsc{IccTA}}~\cite{li2015iccta} leverages \textsc{IC3}~\cite{octeau2015composite} to identify ICC methods and theirs parameters, and then instruments the app by matching and connecting ICC methods with their target components.
The identification of ICC methods and the instrumentation part rely on a list of ICC methods that only contain \emph{well documented ICC methods}.
By considering additional ICC methods (i.e., AICC methods), our tool complements a tool such as \iccta.
In the same way, \textbf{\textsc{DroidSafe}}~\cite{gordon2015information} transforms ICC calls into appropriate method calls to the destination component.
Likewise, \textsc{IccTA}, the ICC methods considered by  \textsc{DroidSafe} are only the well-documented ICC methods. As a result both \textsc{DroidSafe} and \textsc{IccTA} share the same limitation, i.e., they miss the AICC methods. 
Unlike the previously described tools, \textbf{\textsc{Amandroid}}~\cite{wei2014amandroid} constructs an inter-component data flow graph (IDFG) and a data dependence graph (DDG) in which it can run its analysis.
Again, it only considers \emph{documented ICC methods} manipulating \texttt{Intents}.

Other tools leverage ICC links to detect malicious apps. 
\textbf{\textsc{ICCDetector}}~\cite{xu2016iccdetector}, for instance, uses Machine Learning (ML) to detect Android malware. 
The ML model is built by using ICC-related features extracted with \textsc{EPICC}~\cite{octeau2013effective}.
As it relies on \textsc{EPICC} to extract ICC features, it is dependent on \textsc{EPICC} for the considered ICC methods.
Yet, \textsc{EPICC}, just as \textsc{IC3} only considers \emph{documented ICC methods} for inter-component communication.
In the same way, Li \& al.~\cite{li2015potential} set up a ML approach for detecting malicious applications.
The feature set used is based on Potential Component Leaks (PCL) in Android apps.
PCLs are defined using components as entry and/or exit points.
Again, they consider traditional ICC methods as exit points for transferring data through components.

\textbf{\textsc{ICCMATT}}~\cite{jha2015modeling} aims at conceptually modeling ICC in Android apps to generate test cases.
The purpose is to identify components vulnerable to malicious data injection and privacy leaks.
The approach of the researchers takes into account \texttt{PendingIntent} objects, but only at the conceptual level.
They describe them as \texttt{Intent} wrappers able to be shared between components, mainly used in notifications and/or alarm services as we have seen throughout this paper.
They do not directly refer to methods for performing inter-component communication atypically with \texttt{PendingIntent} objects.

In the same way, Enck et al. ~\cite{enck2009understanding} describe the overall functioning of \texttt{PendingIntents} for integration with third-party applications.
Nevertheless, they do not explain, as in~\cite{gross2018pianalyzer}, the security threats that it poses as well as the difficulty it induces for ICC modeling in static analyzers.
PiAnalyzer~\cite{gross2018pianalyzer} models specific vulnerabilities where other apps can intercept broadcasted PendingIntents. In contrast, \raicc{} generically models ICC links where PendingIntent (as well as IntentSender) are involved.
The goals of PiAnalyzer and \raicc{} are thus different. Hence, \raicc{} was not compared to PiAnalyzer in this study.

Besides static analysis approaches, dynamic analysis solutions have also been studied for the detection of ICC data leaks.
For example, \textbf{\textsc{CopperDroid}}~\cite{tam2015copperdroid} is able to reconstruct the app behavior by observing interactions between the app and the underlying Linux system.
\textbf{\textsc{TaintDroid}}~\cite{enck2014taintdroid} dynamically tracks sensitive information with a modified Dalvik virtual machine.
Monitoring the behavior of an Android app is also popular in dynamic data leak detection~\cite{backes2013appguard, hornyack2011these, xu2012aurasium, bartel2012improving}.
Depending on the taint policy set up for propagating tainted data, a dynamic analysis could consider and therefore detect atypical ICC data leaks.
Nonetheless, precise methods exist~\cite{sarwar2013effectiveness} for bypassing taint-tracking, leading to false-negatives as well as more general approaches for tackling ICC-related security issues~\cite{10.1145/3238147.3238207, 10.1145/3106237.3106286}.

%% file: sections/conclusion.tex
\section{Conclusion}
\label{conclusion}

We addressed the challenge of precisely modeling inter-component communication in Android apps. After empirically showing that
 Android apps can leverage atypical ways for performing ICC, we discuss the implications for state of the art ICC modeling-based analysis.
 We contribute towards using methods not primarily made for this purpose.
We have developed  and open-sourced \raicc{}, which reveals AICC methods and further resolves them into standard ICC through instrumentation. We demonstrate that \raicc{} can boost existing analyzers such as \textsc{Amandroid} and \iccta{}, enabling them to substantially increase their data leak detection rates.

%% file: sections/data_availability.tex
\section{Data Availability}
\label{data_availability}

We provide a complete set of artefacts for reproducibility purposes. In particular, we provide the datasets used in our study, third-party tools used as well as scripts to run them. We also provide the entire source code of our tool \raicc.
All artifacts are available online at:
\begin{center}
    \url{https://github.com/Trustworthy-Software/RAICC}
\end{center}

%% file: sections/acknowledgment.tex
\section{Acknowledgment}
\label{acknowledgment}
This work was partly supported 
(1) by the Luxembourg National Research Fund (FNR), under projects CHARACTERIZE C17/IS/11693861, ONNIVA 12696663, and the AFR grant 14596679, 
(2) by the SPARTA project, which has received funding from the European Union's Horizon 2020 research and innovation program under grant agreement No 830892, 
and (3) by the Luxembourg Ministry of Foreign and European Affairs through their Digital4Development (D4D) portfolio under project LuxWAyS.